\documentclass{llncs}
\usepackage{a4,graphicx,fancyhdr, vmargin}
\usepackage{url}
 
\begin{document}
\title{Privacy risk from synthetic data: practical proposals}

  \author{Gillian M Raab\inst{1}}
  \institute{Scottish Centre for Administrative Data Research, University of Edinburgh, Scotland,UK, gillian.raab@ed.ac.uk}
\maketitle
\begin{abstract}
This paper proposes and compares measures of identity and attribute disclosure risk for synthetic data.  Data custodians can use the methods proposed here to inform the decision as to whether to release synthetic versions of confidential data.  Different measures are evaluated on two data sets. Insight into the measures is obtained by examining the details of the records identified as posing a disclosure risk. This leads to methods to identify, and possibly exclude, apparently risky records where the identification or attribution would be expected by someone with background knowledge of the data. The methods described are available as part of the \textbf{synthpop} package for \textbf{R}.  
\end{abstract}

\section{Introduction}
It is now thirty years since the first proposals  were made to use synthetic data (SD) as a privacy enhancing technology (PET) \cite{Rubin1993,Little1993}. These were concerned with  using data synthesis to make confidential versions of microdata available to analysts, while reducing the risk of breaching the privacy of individuals contributing records to the original ``Ground Truth" (GT) data used to create the SD. In recent years  data synthesis has been used in many other contexts, such as anonymizing images that might identify people, or concealing geographic locations from mobile phone transaction data; see \cite{WagnerIsabel2018TPMA} for examples. This paper is restricted to the use of SD to create confidential microdata, often by national statistics agencies (NSAs), as described in a recent UNECE report \cite{starter}. Two recent reviews of the field \cite{DreHae2024,reiter2023} have highlighted the need for  practical privacy metrics to evaluate the disclosure risk (DR) from the release of SD. This is in contrast to the variety of measures of utility available to compare SD with GT\cite{Voas_Williamson_2001,WooProp,KarrAmStat,Snokerss,significance,BowenSnoke2021,raab2021}.

For the past ten years we\footnote{The team was led by Beata Nowok and also included Chris Dibben and the author.} have been developing the \textbf{synthpop} package for creating synthetic microdata. We have mainly implemented methods proposed by others to make them available for use on real problems. The package provides a variety of methods for creating SD and includes tools to allow the person creating the SD  to check that analyses based on the SD  will mirror what would be found from the GT. These can be broadly defined as measures of utility. They include measures that compare results from particular analyses (specific utility) and overall measures that compare the whole distributions (general utility)\cite{Snokerss}.

It has been argued that SD, based only on models, has no records that can be associated with identifiable individuals and is thus not personal data  as defined by GDPR\footnote{See \url{https://commission.europa.eu/law/law-topic/data-protection/data-protection-eu_en}, accessed 19/5/2024 In the UK this is now incorporated within the Data Protection Act 2018}. Most people would now dispute this because it may be possible to infer, from the SD alone, the characteristics of individuals in the GT used in its creation,  This view has been expressed recently by the European Data Protection Supervisor\footnote{See \url{https://edps.europa.eu/press-publications/publications/techsonar/synthetic-data_en} Accessed 20/5/2024}.  Ideally the SD should 
reproduce the relationships between variables that are features of the population that the GT data represent, but obfuscate the random variation about this model; i.e. reproduce the signal but not the noise. There is no agreed definition of what methods can be used to create SD and some models may give too  good a fit to the GT so that the noise can be identified in the SD. Data synthesis carried out via iterative techniques such as GANs \cite{goodfellow2014generative} may be especially prone to this problem, as can popular tree-based models \cite{dandr2011}, such as the default model in \textbf{synthpop}, if the trees are allowed to grow too large.

This paper is concerned with developing tools that data custodians can use to evaluate the DR of SD. In Section 2 we discuss what information needs to be assumed to define the risks. Then in Section 3 we provide details of the scenario we are proposing to define these risks. Measures of identity and attribute DR are defined and recommendations made as to which should be used. The measure ``replicated uniques", $repU$, is proposed for identity disclosure and ``Disclosive in Synthetic Correcy in  Original" ($DiSCO$) for attribute disclosure. Results from one small example are discussed, and include methods to identify cases where an apparent disclosure might not really be an unexpected risk. Section 4 then presents methods of excluding such records using a larger data set. Section 5 discusses the differences and commonalities between disclosure risk and utility of synthetic data and discusses other work on quantifying DR that has developed a different approach to this \cite{giomi2022anon}. The paper concludes with practical recommendations and suggestions for future steps.

\section{Practical considerations.}
Synthetic data has the potential to widen access to administrative data to those outside the organisation where it was created. But the lack of any way of assessing DR is a major barrier to allowing custodians to sanction its release. Those with the responsibility of securing public data cannot be expected to be reassured as to the safety of releasing SD without a practical demonstration of how the privacy of data subjects is being protected.
When NSAs do release data to the public as either tables or microdata (e.g. the Sample of Anonymised Records provided from the UK Censuses\footnote{see \url{https://www.ons.gov.uk/census/aboutcensus/censusproducts/microdatasamples}}) it will have gone through extensive testing and been subject to statistical disclosure control (SDC) to ensure its safety. Section 5 outlines how these procedures could be adapted to synthetic data.

During 2023 the author was part-funded by Research Data Scotland\footnote{An organisation with the mission ``...to work with researchers, analysts and policymakers to unlock the potential of public sector data for the benefit of public good''; see \url{https://www.researchdata.scot/}.} to develop practical measures that could be used to evaluate the DRs of synthetic data that were being considered for release to researchers. The latest version of the \textbf{synthpop} package for \textbf{R} now includes routines to measure disclosure risk\footnote{This is version 1.8.1 that can be installed from Github at \url{https://github.com/gillian-raab/synthpop}.}. Details of the disclosure routines are described in Raab et al. (2024) \cite{raab2024}.
While evaluating the utility of synthetic data is relatively straightforward, the assessment of DR and its integration into a protocol for producing synthetic data is more difficult because it depends on other factors, as discussed in the anonymization decision-making framework \cite{elliot_anonframe}. These include:
\begin{itemize}
    \item {(a)}{
 the context of the data release: when, how, to whom, under what regulation}
     \item {(b)}{ whether the person with access to the SD believes it to be real}
       \item {(c)}{ what other sources of data are available for records in the GT}
   \item {(d)}{ what the person with access to the SD knows about the GT in general and specific ways}
 
    \item {(e)}{ what information is released about how the SD has been created.}

\end{itemize}
Item (a)  is a decision that must be taken by, or on behalf of, the data holder and DR must be interpreted in this context. Item (b) is a major concern to data holders. Someone with SD might inadvertently allow access to others who might believe it to be real leading to a loss of reputation for the data custodian. Another situation where this could arise  would be if a laptop with SD were lost or stolen\footnote{The author's involvement with SD nearly came to an end when her laptop, with SD  created for a training course, was stolen. Fortunately, she was able to reassure the security staff from the data holders that the SD was fully encrypted as well as being clearly labelled as ``Fake data".}. The disclosure measures presented here assume that the SD are approached it as if it were real. While this may seem a worst-case scenario it may be a useful one, since it leads to DR measures that compare risks from the SD to those from the GT, as we illustrate in the examples below. Items (c) and (d) are clearly important. In the next section we explain how they affect the disclosure measures. Item (e) is specific to the means of creating SD. It may include general information about the type of method used in the creation of the SD, as well as specific methods. The latter is much more likely to lead to a privacy violation, especially if details of the model are released to a sophisticated intruder \cite{ShokriReza2021OtPR}. This is an area that would benefit from further investigation. Meanwhile, it would be prudent of data holders not to release details of their synthesis models. 
 
 \section{Scenario and definitions}\label{sec:sc_def_not}
\subsection{Setting the scene}\label{subsec:scen}
These disclosure measures are intended to assess what a person who only has access 
to the SD can infer about known individuals who are present in the GT. We use the term ``intruder" for such a person, though no malicious intent is implied. The intruder is assumed to have information for one or more individuals about the value of certain key variables (quasi-identifiers) that are present in the GT. The identification of quasi-identifiers is an important aspect of DR assessment and a data holder may have to update decisions about this if new data sources are accessed that allow people to discover information about known individuals, e.g. scraping information from the web. The intruder first attempts to see if the individual with these quasi-identifiers is present (identity disclosure), and then to determine the value of other, potentially sensitive, items in the data file that we refer to as targets (attribute disclosure). We are assuming a worst-case scenario where the intruder believes 
they are querying the original data.\footnote{This may not be too unrealistic if the data are made available inadvertently, or if the intruder thinks that efforts to label the SD as e.g. ``Fake Data" are thought to be  just a cover up.}
Disclosure measures from the GT are each compared to similar measures for someone with access only to the SD. The difference between these two measures (e.g. $Dorig$ - $DiSCO$, for the preferred measures described below) is a measure of the disclosure protection afforded by the synthesis.

Here we introduce the measures with an example. Formal definitions with notation and formulae are in Appendix 1.
The first step in evaluating DR, as described here, is to identify a set of keys that might be expected to be known to an intruder. These keys are then combined to form a combination of quasi-identifiers that we designate as $q$. For example, if we have hospital records we might define age, sex date and hospital as keys and this would give a $q$ with levels such as ``\texttt{78 | M | 1/1/2024 | WG}" for a 78 year old man admitted to hospital WG on 1/1/2024.

\subsection{Identity disclosure measures}\label{subsec:ident} 
The concept of k-anonymity is central to identity disclosure for microdata. First proposed in 1998 \cite{kanon1}, it is discussed fully in \cite{elliot_anonframe}. A table is k-anonymous with respect to a set of keys if a record cannot be distinguished from at least $k-1$ others.  Based on this idea, the percentage of records for which the keys identify just one individual (i.e. that breach 2-anonymity or, more simply, are unique in the data) give identity disclosure measures. 
Tables of $q$ values are produced from the GT and the SD. $UiO$ and $UiS$ (Unique in Original and Unique in Synthetic) are the percentages of records in the GT and in the SD with unique values of $q$. An intruder checking out a record for their known set of keys will look for it in the SD. Some records in $UiO$
will not be in the SD and $UiOiS$ (Unique in Original in Synthetic) gives the percentage that would be found. These records are then checked for uniqueness in the
SD, giving $repU$ (replicated uniques) as the percentage of unique GT records that are also unique in the SD.

The percentage $repU$ has been used as a disclosure measure to evaluate SD by \cite{jackson_rss} and by \cite{raab22} \footnote{Jackson et al. in \cite{jackson_rss} argue that the denominator for $repU$ should be the number of records in SD,  rather than those in GT. This is inappropriate because our scenario is to consider the risk to the GT data.}.
Replicated uniques are used in \textbf{synthpop} as part of the statistical disclosure control function, \texttt{sdc}, that includes the option of reducing DR by removing them from the SD. Nowok et al. \cite{nowok_repu} have evaluated this and give an example where this process has very little effect on utility.

\subsection{Attribute disclosure measures}\label{subsec:attrib} 
Attribute disclosure uses the same composite identifier $q$, from the keys, and uses it to identify the level of a target $t$.  The attribute disclosure measure, $DCAP$, was proposed by Elliot \cite{elliot2014SYLLS} and has been used 
in other evaluations of SD \cite{taub_PSD2018,lotte}. $DCAP$ is calculated as the average percentage of records with $t$ correctly predicted from $q$ in the SD.  Its use was proposed for the Synthetic Data Challenge\footnote{This took place at the Newton Institute programme on Data Linkage and Anonymisation, 2016 see \url{https://www.newton.ac.uk/event/dla/}} where teams created SD sets  based on the 2001 Census of Scotland.  In response to criticisms by team members that this measure might be a measure of utility rather than privacy, the measure was adapted to $TCAP$ by restricting disclosure to cases when 
$q$ will predict a unique value of $t$ in the SD - i.e. records with an l-diversity of 1 \cite{ldiv}. $TCAP$ was used in the final report of the data challenge \cite{taub_PSD2018} and in other evaluations of SD \cite{chenUNECE2019,little2022}. It is close to the $DiSCO$ measure that we propose below. Appendix 1 gives formal definitions of these measures.

We approach DR from the point of view of an intruder with access to the SD
and to keys forming $q$ for one or more individuals in the original data.

Modelling what an intruder might do, we calculate the following measures, each of which is a proportion of the original records:
\begin{itemize}
\item{For a given $q$ in  GT search for the $q$ value in the SD. The proportion found becomes $iS$ (in Synthetic)}
\item{Check if all records with the same $q$ in SD have the same level of the target $t$. The proportion passing this further test becomes $DiS$ (Disclosive in Synthetic).}
\item{Then check if these apparent disclosures corresponds to the value of $t$ in the GT. The proportion of GT records for which this is true
becomes $DiSCO$ Disclosive in Synthetic Correct in Original.}
\end{itemize}
Note that records contributing to $DiSCO$ may not be disclosive in the original data;  this information would not be available to the intruder. A further measure $DiSDiO$ (Disclosive in Synthetic Disclosive in Original) restricts the score to those also disclosive in the GT. These measures are defined formally in Appendix 1. 
\subsection{A simple example \label{subsec:simpexamp}}
To illustrate our proposed measures we use a data set about quality of life in Poland (SD2011), available as part of the \textbf{synthpop} package. The target is a score for depression (depress) and $q$ is created from variables ``sex", ``age", ``region" and ``placesize". The depression score has 23 levels and the $q$ formed from these keys has 3,459 levels in the GT but this increases to 4,475 when GT and SD are combined. Almost 50\% of the SD records have unique levels of $q$ and this reduces to 15\% for $repU$.

Five synthetic data sets are created and Table 1 illustrates the attribute disclosure  measures for the target  ``depress" for each of these.  From the GT data 53.3\% ($Dorig$) of records could identify the level of ``depress" with certainty. Turning to the synthetic data we start by seeking the $q$ values from the GT in the SD and 64\%  ($iS$) are found.  The requirement for a record to be disclosive in
the SD reduces this to 33\%  ($DiS$) and again requiring the disclosure to be correct reduces this to around 9\%  ($DiSCO$). This rate reduces further to around 6\%, by restricting to records that were also disclosive in the GT ($DiSDiO$).  

The final column in Table 1 gives the values for $DCAP$, the average percentage of records correctly predicted, but not restricted to those predicted with certainty. As expected the rates are higher than for $DiSCO$. The corresponding rate from the GT ($CAPd$). was 74\% 

\begin{table}[h]
\centering
\begin{tabular}{rrrrrrrcrr}
  \hline
 & $~~~iS$ & $~~~~~DiS$ & $~~~DiSCO$ & $~~~DiSDiO$ & max\_denom & mean\_denom & ~~~~~~~~~~~~~&   $DCAP$ \\ 
  \hline
1 & 64.90 & 34.18 & 9.54 & 6.14 & 3.00 & 1.16& ~~~~~~~& 16.20  \\ 
  2 & 64.00 & 32.50 & 10.26 & 6.78 & 4.00 & 1.19& ~~~~~~~& 17.45  \\ 
  3 & 64.02 & 32.14 & 9.10 & 5.92 & 4.00 & 1.19& ~~~~~~~& 15.92  \\ 
  4 & 63.88 & 33.38 & 9.20 & 5.52 & 4.00 & 1.21& ~~~~~~~& 16.12  \\ 
  5 & 63.44 & 31.46 & 9.34 & 5.80 & 4.00 & 1.23& ~~~~~~~& 16.38  \\ 
   \hline
   \\
\end{tabular}
\caption{\label{tab:t1} Attribute disclosure measures for 5 syntheses of the SD2011 data with target ``depress" identified from keys ``sex", ``age", ``region" and ``placesize" } 
\end{table}

The $Dorig$ and $DiSCO$ measures are not restricted to disclosures that are identified from unique records for a $q$  value and a level of the target in either the GT or the SD. For each disclosive record, the number of GT records with the same  level of the target and the same $q$ as the disclosive records, is the denominator that applies to that record. The columns \texttt{max denom} and \texttt{mean denom} refer to the denominators in the GT that contribute to the $DiSCO$  measures. We can see from the mean that here the majority of disclosive records had unique key combinations in the GT, and the maxima was 3 for the first synthesis and 4 for the others. The disclosure functions in \textbf{synthpop} use information on large denominators to check for a key in $q$ that is highly predictive of one level of the target. When such a relationship is identified it causes a two-way check to be triggered and details printed.
A target where one level accounts for a very high proportion of the disclosures can also lead to high levels of $DiSCO$ for similar reasons and this may be flagged as a one-way check.

 Results for $DiSCO$ as an attribute disclosure measure for the SD and $Dorig$ for the GT are shown in Figure 1, for five targets from the example  discussed above. As well as \texttt{depress}, the other variables are  \texttt{ls}, a score for life satisfaction,  \texttt{workab}, the intention to work abroad, \texttt{marital},  marital status and \texttt{income} income with 8, 3, 9 and 407 distinct values respectively\footnote{including missing value categories}.   The plot is ordered by the disclosure measure for the SD.
 
\begin{figure}[h]
	\centering
	\includegraphics[width=0.98\textwidth]{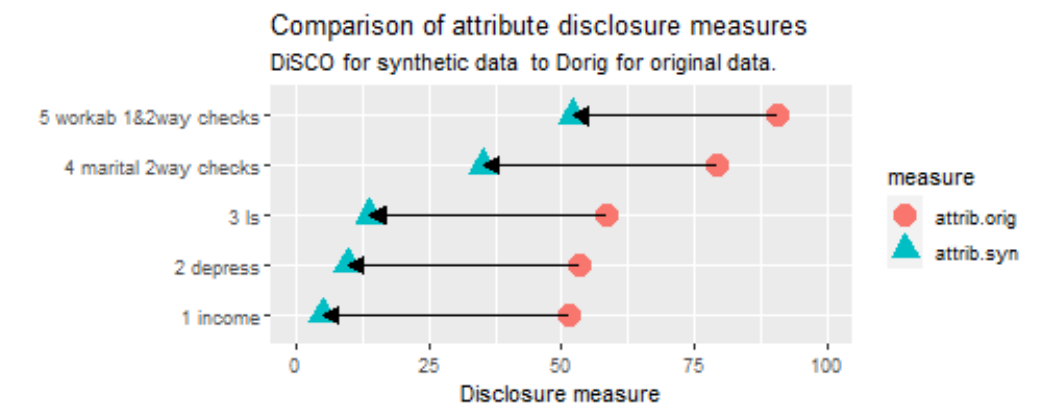}
	\caption{\label{f1}  Attribute disclosure measures $Dorig$ and $DiSCO$ for 5 targets from synthesis of the SD2011 data identified from keys ``sex", ``age", ``region" and ``placesize" } 
\end{figure}

\begin{figure}[h]
	\centering
	\includegraphics[width=0.98\textwidth]{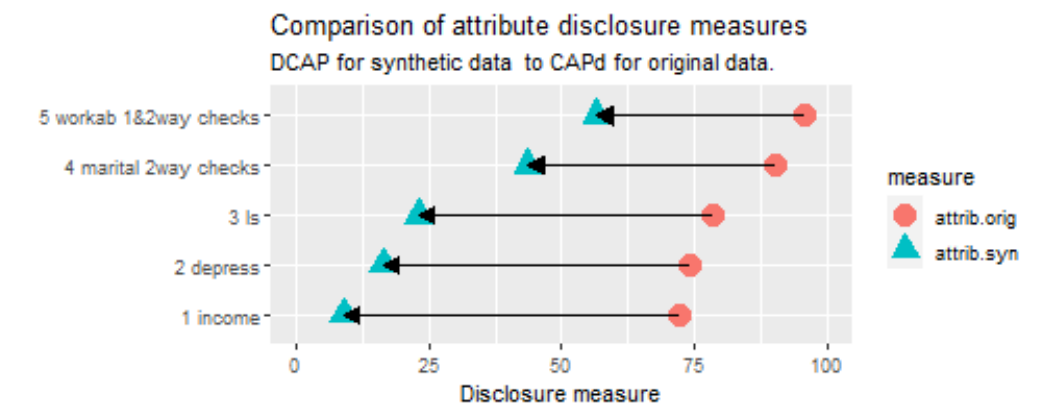}
	\caption{\label{f1b}  Attribute disclosure measures $CAPd$ and $DCAP$ for 5 targets from synthesis of the SD2011 data identified from keys ``sex", ``age", ``region" and ``placesize" } 
\end{figure}

 The top two variables \texttt{workab} and \texttt{marital}  are  flagged as requiring checking; `\texttt{workab} for both one-way and two-way relationships and \texttt{marital} for two-way. Detailed output from the disclosure functions give information on the target levels for one-way checks and the target-key pairs that contribute most to two-way checks. In this example it was the answer  ``NO" for ``workab"  that was flagged by the one-way check;  most survey respondents (89\%) had never worked abroad and 93\% of all disclosive records had this level of the target. For ```marital" several target-key pairs contributed to the two-way check, such as the marital status ``single" for the youngest ages.
 
Figure 2 shows the same analysis as Figure 1, but using $DCAP$ as a disclosure measure. It shows a very similar pattern to Figure 1, although all measures are higher. Most importantly the two variables flagged for one-way or two-way checks show similar disclosure patterns as we saw for the $DiSCO$ measure. These checks, and what to do about them, are discussed further in Section \ref{sec:s3}.

Note that, by default, our disclosure methods treat all variables as if they were categories. Although the \texttt{income} variable may have been an exact value in the original survey, it is rounded in the data that are released so that there are only 406 distinct values in the 5000 records used here. This is typical of all data made available to researchers by NSAs. If 
SD contains more finely-grained continuous data, there are two possible approaches. The first would be to count as disclosure records where the distance between the GT and SD is lower than a threshold. The second is to group the continuous variables into categories that are small enough to make knowledge of them be considered disclosive. The second, simpler, approach is implemented in the\textbf{synthpop} package by allowing the user to specify the number of groups into which each numerical key or target will be categorised. For the target \texttt{income} with 406 distinct values in Figure 1 gave a $DiSCO$ of 4.90\%. Grouping it into 20 categories increased $DiSCO$ to 6.14\%.

\section{\label{sec:s3}Excluding disclosive records that are not risky}
Having identified records where the apparent disclosure is something that would generally be known, one option is to exclude these key-target combinations explicitly from the measures.
For example:
\begin{itemize}
\item{ All records with levels of the target identified as contributing a high proportion of disclosive records can be excluded from all disclosure measures. These can be identified from the one-way checks described in the previous section.}
\item{Selected key-target pairs can be excluded from all measures.
These can be identified from the two-way checks described in the previous section.}
\item{Missing values of some or all of the keys or the targets  can be excluded from leading to a disclosure}
\item{Disclosure measures can be restricted to those disclosive records where the denominators of the disclosive cells for thelevel of $q$ and $t$ are less than or equal to a limit defined by the user. A denominator limit of 1 will restrict disclosure in the SD to records that are unique in the GT for the target $t$ and the conbined identifiers $q$.}
\end{itemize}

\begin{figure}[h]
	\centering
	\includegraphics[width=0.98\textwidth]{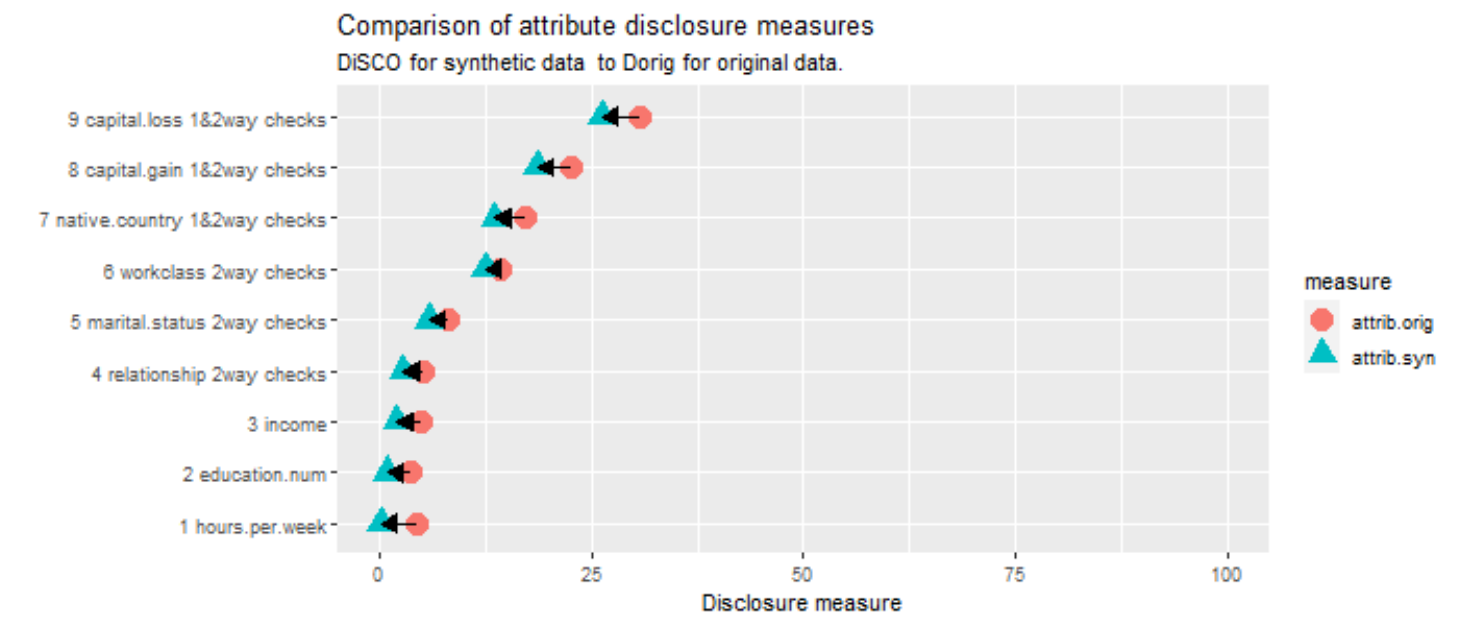}
	\caption{\label{f1b}  Attribute disclosure measures $Dorig$ and $DiSCO$ for 9 targets from synthesis of the Adult data identified from keys ``sex", ``age", ``occupation " and ``race" } 
\end{figure}

To illustrate exclusions, we use the Adult data set from the UCI machine learning repository \cite{UCI} with almost 50 thousand records from the US Census income study. The data set was synthesised with the default method (CART) in \textbf{synthpop}.  Figure 3 sumarises the attribute  DR from the keys \texttt{age, occupation, race and sex} for the other 9 variables in the data\footnote{Although the data set lists 15 variables, two are identical except that one (education) appears as both a factor and a numeric variable and We have excluded the weight, as it is not an analysis variable.}. 
The disclosiveness of the original data is relatively low, compared to the previous example, because of the large sample size and the absence of any geographic identifiers. 
The three variables with the highest attribute disclosure  \texttt{capital.gain, capital.loss and native.country} are flagged to check one-way relationships. Details show that the levels contributing to disclosure are
zero for the first two and \texttt{United-States} for \texttt{native.country}. These levels make up 95\%, 92\% and 89\% of all disclosive records, respectively.

Three other targets are flagged as having disclosive two-way relationships \texttt{workclass, marital, relationship}. Detailed output from the disclosure functions showed that 
the largest contribution to two-way disclosures for \texttt{workclass} was due to this being missing when occupation was missing, although other relationships between these two variables also contributed. For \texttt{marital}, the largest contributions came from the key \texttt{age} with patterns similar to \texttt{marital} in our first example. 

\begin{table}[h]
\centering
\begin{tabular}{|c|rr|rr|rr|rr|rr|}
 \hline
    & \multicolumn{2}{c} {Column 0} & \multicolumn{2}{|c|} {~~~Column 1~~~} & \multicolumn{2}{|c|} {~~~Column 2~~~} & 
   \multicolumn{2}{|c|} {Column 3} &  \multicolumn{2}{|c|} {Column 4} \\
   & \multicolumn{2}{c} {} & \multicolumn{2}{|c|} {One target} & \multicolumn{2}{|c|} {missing} & 
   \multicolumn{2}{|c|} {denominator} &  \multicolumn{2}{|c|} {denominator} \\
  & \multicolumn{2}{c} {No} & \multicolumn{2}{|c|} {level } & \multicolumn{2}{|c|} { values} & 
   \multicolumn{2}{|c|} {limit  1} &  \multicolumn{2}{|c|} {limit 1} \\
 & \multicolumn{2}{c} {exclusions} & \multicolumn{2}{|c|} {excluded} & \multicolumn{2}{|c|} {excluded} & 
   \multicolumn{2}{c} {and a) and b)} &  \multicolumn{2}{|c|} {only} \\
 target & orig~~~ & syn & orig & syn  & orig & syn & orig & syn & orig & syn   \\ 
  \hline
capital.gain  & 22.55 & 18.83 & 0.21 & 0.00 & 0.21 & 0.00 & 0.21 & 0.00 & 2.68 & 1.18 \\ 
  capital.loss  & 30.61 & 26.35 & 0.08 & 0.00 & 0.08 & 0.00 & 0.08 & 0.00 & 2.68 & 1.23 \\ 
  education.& 3.71 & 1.00 & 3.71 & 1.00 & 3.71 & 1.00 & 2.68 & 0.44 & 2.68 & 0.44 \\ 
  hours.per.week & 4.36 & 0.19 & 4.36 & 0.19 & 4.36 & 0.19 & 2.68 & 0.14 & 2.68 & 0.14 \\ 
  income & 4.97 & 2.10 & 4.97 & 2.10 & 3.51 & 1.74 & 1.74 & 0.58 & 2.68 & 0.79 \\ 
  marital.status  & 8.23 & 5.18 & 8.23 & 5.18 & 8.23 & 5.18 & 2.68 & 0.77 & 2.68 & 0.77 \\ 
  native.country & 17.09 & 13.85 & 0.94 & 0.09 & 0.83 & 0.08 & 0.73 & 0.07 & 2.68 & 0.86 \\ 
  relationship  & 5.17 & 2.66 & 5.17 & 2.66 & 5.17 & 2.66 & 2.68 & 0.72 & 2.68 & 0.72 \\ 
  workclass  & 14.27 & 11.71 & 14.27 & 11.71 & 9.14 & 6.91 & 2.45 & 0.81 & 2.68 & 0.93 \\ 
   \hline
   \\
\end{tabular}
 \caption{Attribute disclosure measured by $DiSCO$ from Adult data from the keys \texttt{age, occupation, race and sex} and different exclusion criteria.}
\label{table:1}
\end{table}

Table 2 gives the results of excluding different entries in the tables of $q$ and $t$ from the attribute disclosure measures.
Excluding the levels of the target flagged by one-way checks (capital gain, capital loss and native country) reduces the $DiSCO$ to almost zero for these 3 variables.
Adding the exclusion of missing values reduced the disclosure for \texttt{workclass} and for some other variables a little. 
Adding the restriction to denominators of 1 reduced the disclosure for variables, identified as requiring  two-way checks,  to low 
levels. In the final columns we see that restricting to denominators of 1, by itself, gives levels of disclosure below 1\% for all but the first two variables.
Note that the attribute disclosure of the original data  with denominators of 1 is the same for all targets at 2.68\%. This is the same 
as the \% unique values of $q$ in the GT, because all records with a unique $q$ are disclosive in the GT for all targets.

Which of these columns in Table 2 represents an approach that an NSA should adopt to exclude these one-way and two-way contributions from the attribute disclosure measures? In the case of the Adult data this is a rhetorical question because the data have already been made available to the public. Both column 3 or column 4 give low risks. Column 2 gives low risks except for the three targets identified as having contributions from two-way $tq$ relationships. This could be remedied by excluding the specific $tq$ combinations, but at the cost of more examination of the data. Column 4 offers a quick and easy solution, but perhaps at the price of failing to count some really disclosive records in the DRs.

 The choice of using the easy measure (Column 4) or the more onerous one (modified column 2) may depend on the context of the release, as discussed in Section  2, factor (a). For SD that will be released to trusted researchers, with guarantees that it will not be made public, the easy measure given by the method in Column 4 might be enough. But for SD to be openly shared, the data holder would be expected to carry out more extensive DR assessment.

This paper presents only a very limited  evaluation of the DR measures, using two examples. It is hoped that the availability of the new \texttt{synthpop} tools for DR assessment will increase our understaning of DR, the factors that influence it and how best it should be assessed.

\section{Can we or should we exclude utility from attribute  disclosure risk and does it matter?}
Inevitably, measures of utility and attribute  DR have much in common.  Attribution risk involves the ability to predict one variable from others. This is clearly related to utility. In a comprehensive review of privacy metrics Wagner and Eckhoff \cite{WagnerIsabel2018TPMA} note that authors have suggested that ``privacy metrics are orthogonal to cost and utility metrics". Yet, several of the metrics they propose are have also been cited as general utility metrics \cite{raab2021}.

When disclosure is an aspect of utility, it may still be a real disclosure for the individual concerned, if it correctly predicted a sensitive attribute. The exception would be if it would be the result of a relationships in the data that was already known. Ideally we would have a full specification of the intruder's prior belief about what the relationships in the data might be. With that information we
could assess the additional information that the SD provides compared to the prior\cite{HUReiter2014}. Specification of a prior would be an onerous task, so the simplification we propose in Section 4 is an attempt to identify some of the prior information that might be available for low-dimensional relationships in the data\footnote{While restricting to one-way or two-way relationships seems limited, in practice stratification by segments of the GT such as area can make this more effective.}. It is specific to the knowledge of the intruder. For example, a table of UK occupations by income might identify that all university vice-chancellors had incomes in the highest group. Whether this would be a new DR would depend on what was known about UK academic salaries.

A recent proposal of a framework for quantifying privacy risk, entitled the Anonymeter, has been proposed by Giomi et al. \cite{giomi2022anon} who provide open source code at \url{https://www.anonos.com/products/anonymeter} where they also post a quote from a letter to them from the French data protection authority, the Commission Nationale de l’Informatique et des Libertés (CNIL) as:
\begin{quote}
"The results produced by the tool Anonymeter should be used by the data controller to decide whether the residual risks of re-identification are acceptable or not, and whether the dataset could be considered anonymous."
\end{quote}
They propose methods to assess privacy risk as \textit{singling out}, \textit{linkability} and \textit{inference}, Their measures of \textit{singling out}  and \textit{inference} appear to be identical to the measures $RepU$ and $DCAP$ described here. However they are computed by a simulation rather than from the exact formulae given in Appendix 1. They claim that the Anonymeter provides a method that makes it possible to ``measure how much of an attacker's success is simply due to the \textit{utility} of the SD and how much instead is an indication of \textit{privacy violation}".  They propose is the following:
\begin{itemize}
\item{divide the GT data into two sections as \texttt{Training} and \texttt{Control}. \texttt{Training} is used to create the SD and the disclosiveness is assessed from \texttt{Control} They suggest that \texttt{Control} might be a smaller fraction (e.g. 20\%) so as to maintain the quality of the SD}
\item{create SD based on the Training data}
\item{measure the DR for the Control data from this SD.}
\end{itemize}
The authors argue that the success rate of the Control is a consequence of general inference (i.e. utility) and the extent to which the Training risk exceeds the Control is an indication that information has been leaked from the SD. They propose a measure $R$ which the term a ``specific privacy risk" as
\begin{equation}
R = {{  (r_{train} - r_{control})  }\over{1-r_{control}}},
\end{equation}
where $r_{train}$ and $r_{control}$ are the proportions of disclosive records identified in the control and training data. It is intended to measure the proportion of the DR that excludes disclosure due to utility ($r_{control}$).
A limited comparison of this method has been carried out with the analyses of attribute disclosure from the example in Section 3.4. For the five targets illustrated in Figures 1 and 2, attribute disclosure was calculated using the Anonymeter methodology for a Control with 1000 records (20\% of the GT data) and a Training data set of the remaining  80\%. Results are given in Table 3.

\begin{table}[ht]
\centering
\begin{tabular}{crrrrrrrr}
  \hline
   & \multicolumn{4}{c} {$DiSCO$ } & \multicolumn{4}{c}{$DCAP$} \\
  & ~~~All & ~~~Train & ~~~Ctrl  &~~~~~~ R  & ~~~~~~~~~~~~All & ~~~Train & ~~~Ctrl & ~~~~~~R \\ 
  \hline
income & 4.90 & 5.53 & 1.90 & 0.04 & 8.91 & 9.06 & 2.92 & 0.06 \\ 
depress & 9.54 & 9.45 & 4.20 & 0.05 & 16.39 & 15.98 & 5.80 & 0.11 \\ 
ls & 13.78 & 11.88 & 8.60 & 0.04 & 23.03 & 19.79 & 12.53 & 0.08 \\ 
marital & 35.18 & 34.50 & 21.30 & 0.17 & 43.66 & 41.50 & 24.94 & 0.22 \\ 
workab & 52.10 & 49.28 & 31.30 & 0.26 & 56.56 & 52.92 & 33.31 & 0.29 \\ 
   \hline
\\\\
\end{tabular}

\caption{Attribute  disclosure measures $DiSCO$ and $DCAP$ calculated from the the training and control segments of the GT data with SD created by the training data. The columns ``All" are each measure from the complete data as illustrated in Figures 1 and 2. }
\end{table}
There is some agreement here with the results in Figures 1 and 2. The two targets (marital and workab) that were identified as having large contributions from one-way or two-way relationships are also those identified here as having the largest /boldfont{R} ratios that \cite{giomi2022anon} identify as being the proportion of attribute disclosure due utility.  The attribute disclosure measures for ``workab" after excluding records answering ``NO" as disclosive became 2.5\% for $DiSCO$ and 3.21\% for $DCAP$ are much lower than the  Ctrl values in Table 3. They would give /boldfont{R} ratios of 0.51 and 0.55 and suggests that the anonymeter procedure may not be identifying all of the disclosure that can be explained by utility for this example. The same was true for ``marital" when $tq$ combinations identified from a two-way table were excluded from the disclosure measures. 

Further investigations of the comparison between the anonymeter methods and what is proposed in this paper may  help the understanding of both methods. It is interesting to note the qualification by the CNIL after the quote from their letter above:

\begin{quote}
"The anonymity of a synthetic dataset can only be determined on a case-by-case basis, i.e. for each generated dataset, and should therefore not be assumed from analyses performed on other datasets coming from the same provider or data synthesis tool. 
However, as a solution provider, the company Statice should also provide practical elements describing precisely how to use the tool and interpret the results obtained (such as examples, tutorials, thresholds, etc.)." 
\end{quote}

The metrics introduced in Section 3 have been developed for practical use by data custodians who are deciding on the release of each individual SD. This is not so clear in the anonymeter paper \cite{giomi2022anon}. There is no attempt there to identify sensitive variables, or those variables whose value would be expected to be known for individuals in the data. Results are presented that are averaged over all variables in the GT as targets using random selections from the remaining variables as keys.
It may be that their metrics have some value for evaluating the DR of methods to create SD rather than evaluating individual synthetic data sets.

\section{Integrating utility and DR measures into synthetic data creation}

Before a data holder decides to release SD it needs to be evaluated for Utility and DR. The person creating the SD has the means that allow each of these to be altered. They include the synthesis methods that are used, and also SDC procedures that can be carried out on either the GT or the SD. This will need to be an iterative process where the synthesis methods and the SDC are adapted until the SD is satisfactory with respect to utility and DR. 

These modifications of the SD occur at three stages: 
\begin{itemize}
    \item {(a) Pre-processing the GT before synthesis.}
    \item{(b) Changing the methods used to create the SD.}
    \item{(c) Post-processing the SD before it is released.}
\end{itemize}
Techniques that can be used at (a) and (c) include recoding variables with small categories, grouping or smoothing numeric variables and the removal of records that are unique in the GT data or are replicated uniques in SD, as well as other techniques from Statistical disclosure control\cite{sdcbook}.

Under item (b) many different methods can be used to create SD and each may have parameters that can be tuned to influence utility or DR. The default method in \textbf{CART} can be adjusted to improve utility via parameters like the sequence of the conditional distributions and the restriction of the predictors used for each conditional model; see \cite{Raab_Nowok_Dibben_2017} for examples.  An adjustment of the \textbf{CART} method to limit the number of records in the terminal modes of the tree can improve DR.

Using synthesis methods that satisfy differential privacy(DP) have been advocated as a method of reducing the DR of SD. Our methods of assessing DR are, in one sense, the opposite of the formulation of DP. Our proposals are very specific as to what is known about individuals in the data and to knowing which items may be sensitive to being disclosed. Differential privacy, on the contrary, attempts to protect against arbitrary outside knowledge on the part of an intruder. There have been several important criticisms of DP SD: see for example \cite{groundhog} in terms  of its lack of utility, and even its poor DR. These underline the importance of using independent DR assessment even for DP synthesis.

\section{Discussion and future work}
This paper has shown how using routines to calculate DR measures on real data sets can lead to understanding and improving the DR metrics. It is hoped that more feedback will improve the metrics even more. The author would welcome feedback on the new \textbf{synthpop} disclosure measures either to suggest improvements or (more likely at this stage) to report bugs or other problems.

There is much more work that could be done on factors that influence the DR of synthetic data. One important aspect would be to investigate whether over-fitting the synthesis model will increase DR. An example of a potentially over-fitted model is the use of a saturated model for categorical data from a cross-tabulation of all the variables. This has been proposed by Jackson et al. \cite{jackson_et_al2021} and is implemented as the method \texttt{catall} in \textbf{synthpop}. The disclosure of SD created by \texttt{catall} could be compared to that created using the \texttt{ipf} which uses iterative proportional fitting to a set of defined margins to create the SD.

Further work comparing the metrics presented here with those recommended by Anonymeter may help us to clarify the differences between utility and attribute DR for SD.

\section{Acknowledgement}
Research Data Scotland (\url{https://www.researchdata.scot/}) has funded some of  Gillian Raab's time to carry 
out the research reported here and  to expand the capabilities of the \textbf{synthpop} package\footnote{
from version 1.8-0 available on CRAN at \url{https://CRAN.R-project.org/package=synthpop}
}
to include measures of disclosure control. We also thank the Scottish Centre for Administrative 
Data Research for continue to support the development of the \textbf{synthpop} package since its creation was
supported by the ESRC funded SYLLS poject in 2012-14.
We would also like to thank two anonymous refrees for their helpful comments.

\section*{Appendix 1: Notation and formal definitions}{\label{sec:app1}}
\subsection*{Computational approach and notation}
Before defining the measures of identity and attribute DRs we need to introduce the notation that will be used to calculate them. The first step is to create the quasi-identifiers from the keys for the GT and SD. For the keys used in the example given in Section \ref{subsec:simpexamp} the quasi-identifier that we will designate as $q$ for the first record in the original data is:
\\
\noindent{\texttt{"FEMALE | 57 | Lubuskie | URBAN 100,000-200,000"}}

\noindent{and that for the first record in the synthetic data:}

\noindent{\texttt{"FEMALE | 39 | Zachodnio-pomorskie | URBAN 100,000-200,000''}}.
\\
In order to calculate identity disclosure measures, we need to compare the tables of $q$ from the GT and SD. For attribute disclosure measures we need to cross-tabulate $q$ with each target variable $t$ and compare findings from the SD with what would have been found from the GT. In general, the levels of $q$ and sometimes $t$ in the GT and SD will not be the same. Before creating any tables, we need to define sets of $q$ and $t$ values that give the union of both sets of levels and align the tables so that their indices correspond.

For the GT data $d_{.q}$ is the count of records with the keys corresponding to the levels of $q$ and $d_{tq}$ the count of records with this $q$ and level $t=1,...T$ of the target. The equivalent counts from the synthesised data are designated by $s_{.q}$ and $s_{tq}$. When a member of $q$ is in the GT data but not in the SD, $s_{.q}$ and $s_{tq}$ are all zero. Similarly when a member of $q$ is in the SD but not in the GT, $d_{.q}$ and $d_{tq}$ are all zero. The two tables can be written as shown in Table 4, where the total records in the GT data is $N_d$, made up of $N_{d~only}$ and $N_{d~both}$. The
equivalent totals for the SD are $N_s$, $N_{s~only}$ and $N_{s~both}$.

\begin{centering}
\begin{table}[h]
\begin{tabular}{c|ccc|ccc|ccc|c} 
 & \multicolumn{3}{|c|}{only in original} & \multicolumn{3}{|c|}{in both} & \multicolumn{3}{|c|}{only in synthetic} & Total\\
 \hline
1  & ... & $d_{1q}$ & ...  & ... & $d_{1q}$ & ...  & ... & 0 & ... & $d_{1.}$ \\
 ... & ... & ... & ...  & ... & ... & ...  & ... & ... & ... & ... \\
t  & ... & $d_{tq}$  & ...  & ... & $d_{tq}$ & ...  & ... & 0 & ... & $d_{t.}$ \\
... & ... & ... & ...  & ... & ... & ...  & ... & ,,, & ... & ... \\
T  & ... & $d_{Tq}$ & ...  & ... & $d_{Tq}$ & ...  & ... & 0 & ... & $d_{T.}$ \\
 \hline
Column sums  & & $d_{.q}$ &  & ... & $d_{.q}$ & ...  & ... & 0 & ... & $N_d$ \\
 \hline
Totals  &  & $N_{d\:only}$ &  & & $N_{d~both}$ &  & & 0 &  & $N_d$ \\\\\\

 & \multicolumn{3}{|c|}{only in original} & \multicolumn{3}{|c|}{in both} & \multicolumn{3}{|c|}{only in synthetic} & Total\\
 \hline
1  & ... & 0 & ...  & ... & $s_1q$ & ...  & ... & $s_1q$ & ... & $s_{1.}$ \\
 ... & ... & ... & ...  & ... & ... & ...  & ... & ... & ... & ... \\
t  & ... & 0 & ...  & ... & $s_tq$ & ...  & ... & $s_tq$ & ... & $s_{t.}$ \\
... & ... & ... & ...  & ... & ... & ...  & ... & ... & ... & ... \\
T  & ... & 0 & ...  & ... & $s_Tq$ & ...  & ... & $s_Tq$ & ... & $s_{T.}$ \\
 \hline
Column sums  & ... & 0 & ...  & ... & $s_{.q}$ & ...  & ... & $s_{.q}$ & ... & $N_s$ \\
 \hline
Totals  &  & 0 &  & & $N_{s~both}$ &  &  &  & $N_{s\:only}$ & $N_s$ \\\\

\end{tabular}
\caption{Notation for tables from quasi-identifier ($q$) and target ($t$) from GT (upper table) and SD (lower table).}
\label{table:2}
\end{table}
\end{centering}
\subsection*{Identity disclosure}
 To calculate the \% of unique records in the GT and synthetic data we need:
  \begin{equation}
\%~Unique~in~Original = UiO = 100\sum{(d_{.q} |d_{.q} = 1})/N_d. 
  \end{equation}
  \begin{equation}
  \%~Unique~in~Synthetic = UiS = 100\sum{(s_{.q} |d_{.q} = 1})/N_d.
    \end{equation}
    The intruder has information about the keys for an individual in the GT that they attempt to identify in the SD. They first attempt to find them in the SD, and  the \% found is:
      \begin{equation}
    \%~Unique~in~Original~in~Synthetic = UiOiS = 100 \sum{(d_{.q} = 1 |s_{.q} = 1 \land d_{.q} > 0})/N_d.
      \end{equation}
      Some of these records would not be unique in the SD, restricting to such records gives:
        \begin{equation}
      \%~replicated~Uniques = repU = 100\sum{(s_{.q} |d_{.q} = 1 \land s_{.q} = 1)}/N_d.
      \end{equation}
     \subsection*{Attribute disclosure} 
      To find an attribute from a set of keys, it is necessary to examine the distribution of $s_{tq}$ for groups defined by $q$. We define column proportions for the GT and SD as $pd_{tq} = d_{tq}/d_{.q}$ and for the synthetic as $ps_{tq} = s_{tq}/s_{.q}$.

The measure  $DCAP$ is the percentage of $t$ correctly predicted in the SD:  This gives
\begin{equation}
  \nonumber DCAP = 100 \sum\limits_{tq}{(ps_{tq}}d_{tq})/N_d
\end{equation}

As a comparator for $DCAP$ we need the \% that would be predicted with someone with access to the GT, giving $CAPd$:
\begin{equation}
  \nonumber CAPd = 100 \sum\limits_{tq}{(pd_{tq}}d_{tq})/N_d,
\end{equation}

Returning to the scenario described in Section \ref{subsec:scen}, we  first calculate a  measure of attribute disclosure for the GT data that requires that each set of records defined by $q$ has a unique value  of the target. This is an attribute disclosure measure for the GT data :  \textit{\% Disclosive in Original} :
\begin{equation}
 Dorig = 100\sum^q{\sum^t{(d_{tq} |pd_{tq} = 1})}/N_d.
\end{equation}
An intruder with access only to the SD, but with knowledge of $q$ from one or more individuals in the GT, would look them up in the SD. Some of their $q$ levels be key combinations that do not appear in the SD
leaving the proportion that do appear as $iS$ (in in Synthetic) 
\begin{equation}
 iS = 100\sum^q{\sum^t{(d_{tq} | s_{tq}>0} )}/N_d.
\end{equation}
A level of $q$ from a GT record with level $i$ of $t$ may identify any level $j$  of the target as disclosive in the SD giving
\begin{equation}
 DiS = 100\sum^q{\sum^{i=1,..T}\sum^{j=1,..T}{(d_{iq} | ps_{jq} = 1}  )}/N_d.
\end{equation}
Some of these will identify the wrong target.
To exclude these we restrict to records where $i = j$ giving Disclosive in Synthetic Correct Original:
\begin{equation}
 DiSCO = 100\sum^q{\sum^t{(d_{tq} | ps_{tq} = 1} }/N_d.
\end{equation}

Note that $DiSCO$ can include records that are not disclosive in the GT data giving a further 
measure Disclosive in Synthetic and Disclosive in the  Original :
\begin{equation}
DiSDiO = 100\sum^q{\sum^t{(d_{tq} | ps_{tq} = 1 \land pd_{tq} = 1}  ) }/N_d.
  \end{equation}
As we comment above the intruder would not be able to tell if records were 
identified as $DiSDiO$ rather than $DiSCO$, so we prefer the latter measure.
However, the intruder can identify when the apparently disclosive record
is not unique in the SD. 
 This restriction can be imposed by requiring the denominator in the SD not to exceed a 1, as described
 and discussed in Section 4.

The measures $DCAP$ and $DiSCO$ computed by \textbf{synthpop} use the total number of records in the GT as a denominator. This differs from the measures  used by  \cite{elliot2014SYLLS} and \cite{taub_PSD2018}.  Their $DCAP$ is the same as equation 6, except that it uses as  denominator $N_{d~both}$, the count of records in the GT that have $q$ values represented in the SD.
This denominator is also used to calculate $TCAP$ in \cite{little2022}, thus differentiating it from $DiSCO$, giving 
\begin{equation}
 TCAP = 100\sum^q{\sum^t{(d_{tq} | ps_{tq} = 1} })/N_{d~both}.
\end{equation} 

Both $DCAP$ and $TCAP$ can be scaled with respect to the disclosure that would be found for someone who had access to the marginal distribution of the target. 
 The intruder  guesses the level for each target according to the relative frequencies $pd_{t.}$ in the GT. Averaging this over all observations gives
\begin{equation}
  \nonumber baseCAPd = \sum{(pd_{t.})^2}.
\end{equation}
The $DCAP$ or $TCAP$ score can be expressed  by scaling  from 1 to $baseCAPd$ \cite{little2022,lotte}, although this can result in some negative values.

\section*{Appendix 2: \textbf{R} code for examples analysed in Sections 3 and 4}
\begin{verbatim}
################### R code ########################################
devtools::install_github("Gillian-Raab/synthpop",build_vignettes = TRUE)
library(synthpop)
rm(list=ls())
ods <- SD2011[, c("sex", "age", "region","placesize","depress",
                  "income","ls","marital" , "workab")]
###------------------------------- Table 1 ------------------------------
s1 <- syn(ods, seed = 8564, cont.na = list(income = -8))
s5 <- syn(ods, seed = 8564, m = 5, cont.na = list(income = -8))
t5 <- disclosure(s5, ods, print.flag = TRUE, target = "depress",
       keys = c("sex", "age", "region", "placesize"))
t5
###-------------------------------- Figures 1 and 2 -------------------------------
tt1 <- multi.disclosure(s1, ods, keys = c("sex", "age", "region", "placesize"))
tt1
tt1b <- multi.disclosure(s1, ods, attrib.meas ="DCAP", keys = c("sex", "age", "region", "placesize"))
tt1b
## grouping
disclosure(s1, ods, target = "income", keys = c("sex", "age", "region", "placesize"))
disclosure(s1, ods, target = "income", ngroups_target = 20, keys = c("sex", "age", "region", "placesize"))
##----------------------------- ADult data set anals----------------------
library(arules)
data(AdultUCI)
##syn.AdultUCI <- syn(AdultUCI) gives warning about "-" in variable names
names(AdultUCI) <- gsub("-",".", names(AdultUCI))
myAdult <- AdultUCI
for (i in c(5,15)) myAdult[,i] <- factor(as.character(myAdult[,i])) ## logical and ordered factor changed to factor
myAdult <- myAdult[,-c(3:4)] ## drop fnlwgt and one of education
# order changed to put education.num (now a factor) at the end
system.time(
  syn1.myAdult<- syn(myAdult, cont.na = list(capital.gain = 99999, hours.per.week =99), 
                     method = c("sample",rep("ctree", 12)), visit.sequence = c(1,2,4:13,3) ,models = TRUE)
)
summary.disc1 <- multi.disclosure(syn1.myAdult, myAdult,
                                    key = c("age","sex","occupation","race"))
summary.disc1

###----------------------- table 2----------------------
tab1 <- summary.disc1$attrib.table[,1:2]
dimnames(tab1)[[1]] <- substring(dimnames(tab1)[[1]],3)
dimnames(tab1)[[1]]
tab1 <- tab1[order(dimnames(tab1)[[1]]),]; tab1
targs <-names(myAdult)[!names(myAdult) %in% c("age","sex","occupation","race")]
summary.disc1_nottarg <- multi.disclosure(syn1.myAdult, myAdult,
                                            not.targetslev= c("","","","","0","0","","United-States",""),
                                            key = c("age","sex","occupation","race"))
tab2 <- summary.disc1_nottarg$attrib.table[,1:2]
dimnames(tab2)[[1]] <- substring(dimnames(tab2)[[1]],3)
tab2 <- tab2[order(dimnames(tab2)[[1]]),]; tab2
summary.disc1NA <- multi.disclosure(syn1.myAdult, myAdult,usetargetsNA = FALSE,
                                      not.targetslev= c("","","","","0","0","","United-States",""),
                                      key = c("age","sex","occupation","race"))
tab3 <- summary.disc1NA$attrib.table[,1:2]
dimnames(tab3)[[1]] <- substring(dimnames(tab3)[[1]],3)
tab3 <- tab3[order(dimnames(tab3)[[1]]),]; tab3
summary.disc1NAd1 <- multi.disclosure(syn1.myAdult, myAdult,usetargetsNA = FALSE,
                                        denom_lim =1,  exclude_ov_denom_lim = TRUE,  not.targetslev= c("","","","","0","0","","United-States",""),
                                        key = c("age","sex","occupation","race"))
tab4 <- summary.disc1NAd1$attrib.table[,1:2]
dimnames(tab4)[[1]] <- substring(dimnames(tab4)[[1]],3)
tab4 <- tab4[order(dimnames(tab4)[[1]]),]; tab4
summary.d1 <- multi.disclosure(syn1.myAdult, myAdult,  denom_lim =1, 
                                 exclude_ov_denom_lim=TRUE,                        key = c("age","sex","occupation","race"))
tab5 <- summary.d1$attrib.table[,1:2]
dimnames(tab5)[[1]] <- substring(dimnames(tab5)[[1]],3)
tab5 <- tab5[order(dimnames(tab5)[[1]]),]; tab5
alltab <- cbind(tab1,tab2,tab3, tab4, tab5)
round(alltab,2)
###################################################################
###------------ anonymeter comparisons Table 3------------------------------
## first from figs 1 and 2
tt1$attrib.table[,2]  ### for DiSCO
tt1b$attrib.table[,2] ### for DCAP

ntrain <- round(dim(ods)[1]*.8)
set.seed(6787)
train.ind <- sample(1:dim(ods)[1],ntrain)
train <- ods[train.ind,] 
control <- ods[-train.ind,]
syntrain <- syn(train, seed = 8564, cont.na = list(income = -8)) 
train.disc <- multi.disclosure(syntrain,train, keys = c("sex", "age", "region", "placesize"))
train.disc$attrib.table[,1:2]
control.disc <- multi.disclosure(syntrain, control, keys = c("sex", "age", "region", "placesize"))
control.disc$attrib.table[,1:2]
train.disc.DCAP <- multi.disclosure(syntrain,train, keys = c("sex", "age", "region", "placesize"),
attrib.meas = "DCAP")
train.disc.DCAP[,1:2]
control.disc.DCAP <- multi.disclosure(syntrain, control, keys = c("sex", "age", "region", "placesize"),
attrib.meas = "DCAP")
control.disc.DCAP$attrib.table[,1:2]


table3 <- cbind(tt1$attrib.table[,2],
train.disc$attrib.table[,2],control.disc$attrib.table[,2])
table3 <- cbind(table3[,1:3],
(table3[,2]-table3[,3])/(100-table3[,3]))
table3.DCAP <- cbind( tt1b$attrib.table[,2],train.disc.DCAP$attrib.table[,2],control.disc.DCAP$attrib.table[,2])
table3.DCAP <- cbind(table3.DCAP[,1:3], 
(table3.DCAP[,2]-table3.DCAP[,3])/(100-table3.DCAP[,3]))
table3 <- cbind(table3, table3.DCAP)
dimnames(table3)[[1]] <-dimnames(control.disc.DCAP$attrib.table)[[1]]
dimnames(table3)[[2]] <- rep(c("All","train","control","R"),2)
round(table3,2)
###------------workab excluding NO answers from  disclosure ------------------

tt1x <- disclosure(s1, ods, attrib.meas ="DCAP", 
not.targetlev= "NO",target="workab",keys = c("sex", "age", "region", "placesize"))
tt1x$attrib
(52.10-2.46)/(100-2.46)  ## 0.51 DiSCO

tt1bx <- disclosure(s1, ods, attrib.meas ="DCAP", 
not.targetlev= "NO",target="workab",keys = c("sex", "age", "region", "placesize"))
tt1bx$allCAPs
(56.65 - 3.21)/(100 - 3.21) ## 0.55 DCAP###------------workab excluding NO answers from  disclosute ------------------

tt1x <- disclosure(s1, ods, attrib.meas ="DCAP", 
not.targetlev= "NO",target="workab",keys = c("sex", "age", "region", "placesize"))
tt1x$attrib
(52.10-2.46)/(100-2.46)  ## 0.51 DiSCO

tt1bx <- disclosure(s1, ods, attrib.meas ="DCAP", 
not.targetlev= "NO",target="workab",keys = c("sex", "age", "region", "placesize"))
tt1bx$allCAPs
(56.65 - 3.21)/(100 - 3.21) ## 0.55 DCAP
\end{verbatim}
\end{document}